\documentclass{emulateapj} %
\usepackage{graphicx}

\def\arcsec{$\,^{\prime\prime}$~}
\def\arcmin{$\,^\prime$~}

\newcommand{\be}{\begin{equation}}
\newcommand{\bel}[1]{\begin{equation}\label{eq:#1}}
\newcommand{\ee}{\end{equation}}
\newcommand{\bd}{\begin{displaymath}} 
\newcommand{\ed}{\end{displaymath}}   
\newcommand{\bea}{\begin{eqnarray}}
\newcommand{\beal}[1]{\begin{eqnarray}\label{eq:#1}}
\newcommand{\eea}{\end{eqnarray}}

\newcommand{\eqref}[1]{\ref{eq:#1}}

\newcommand{\lsim }{{\lower0.8ex\hbox{$\buildrel <\over\sim$}}}
\newcommand{\gsim }{{\lower0.8ex\hbox{$\buildrel >\over\sim$}}}

\def\Chandra{${\it Chandra}$}

\def\simge{\mathrel{%
   \rlap{\raise 0.511ex \hbox{$>$}}{\lower 0.511ex \hbox{$\sim$}}}}
\def\simle{\mathrel{
   \rlap{\raise 0.511ex \hbox{$<$}}{\lower 0.511ex \hbox{$\sim$}}}}

\newcommand{\Msun}{\ifmmode {M_{\odot}}\else${M_{\odot}}$\fi}
\newcommand{\Lsun}{\ifmmode {L_{\odot}}\else${L_{\odot}}$\fi}
\newcommand{\Rsun}{\ifmmode {R_{\odot}}\else${R_{\odot}}$\fi}

\shorttitle{X-ray Spectra of Unusual CV in M3}
\shortauthors{Stacey et al.}

\begin{document}
\title{Transient Extremely Soft X-ray Emission from the Unusually Bright CV in the Globular Cluster M3: a New CV X-ray Luminosity Record?}  

\author{W.~S. Stacey\altaffilmark{1}, C.~O. Heinke\altaffilmark{1,2,3}, R.~F. Elsner\altaffilmark{4}, P.~D. Edmonds\altaffilmark{5}, M.~C. Weisskopf\altaffilmark{4}, J.~E. Grindlay\altaffilmark{5}}

\altaffiltext{1}{University of Alberta Physics Dept., 11322-89 Avenue, Edmonton, AB T6G 2G7}
\altaffiltext{2}{Ingenuity New Faculty}
\altaffiltext{3}{To whom correspondence should be addressed; heinke@ualberta.ca}
\altaffiltext{4}{NASA Marshall Space Flight Center, VP62, Huntsville, AL 35812}
\altaffiltext{5}{Harvard-Smithsonian Center for Astrophysics, 60 Garden Street, Cambridge, MA 02138, USA}


\begin{abstract}
We observed the accreting white dwarf 1E1339.8+2837 (1E1339) in the globular cluster M3 in Nov. 2003, May 2004 and Jan. 2005, using the \Chandra\ ACIS-S detector. 
The source was observed in 1992 to possess traits of a supersoft X-ray source (SSS), with a 0.1-2.4 keV luminosity as large as $2 \times10^{35}$ erg s$^{-1}$, after which time the source's luminosity fell by roughly two orders of magnitude, adopting a hard X-ray spectrum more typical of CVs. 
Our observations confirm 1E1339's hard CV-like spectrum, with photon index $\Gamma=1.3\pm0.2$. We found 1E1339 to be highly variable, with a 0.5-10 keV luminosity ranging from $1.4\pm0.3\times10^{34}$ erg s$^{-1}$ to $8.5^{+4.9}_{-4.6}\times10^{32}$ erg s$^{-1}$, with 1E1339's maximum luminosity being perhaps the highest yet recorded for hard X-ray emission onto a white dwarf. In Jan. 2005, 1E1339 displayed substantial low-energy emission below $\sim0.3$ keV.   Although current \Chandra\ responses cannot properly model this emission, its bolometric luminosity appears comparable to or greater than that of the hard spectral component. This raises the possibility that the supersoft X-ray emission seen from 1E1339 in 1992 may have shifted to the far-UV.

\end{abstract}


\maketitle


\section{Introduction}\label{s:intro}

Supersoft X-ray sources (SSS) were first detected by the Einstein observatory and were later established as a prominent X-ray binary class upon the basis of {\it ROSAT} observations \citep{Greiner91}. SSSs are distinguished from other X-ray sources by their soft spectra ({\it kT} ${\sim}$15-80 eV) and high bolometric luminosities, generally ranging from $10^{36}$ erg s$^{-1}$ to $10^{38}$ erg s$^{-1}$.
 These attributes are most suitably explained by a binary system with a white dwarf (WD) primary in which hydrogen-rich matter is burned upon the surface of the WD, accreting at a rate of $\sim$ $10^{-7}$ $M_{\sun}$ year$^{-1}$ \citep{Kahabka97}.  
 
Systems portraying characteristics of SSSs have been considered as possible progenitors of Type Ia supernovae \citep{Rappaport94}. Type Ia supernovae are triggered by one of two pathways: the steady accumulation of matter onto an accreting white dwarf, until it passes the Chandrasekhar limit \citep{Whelan73}, or the merger of two white dwarfs \citep{Webbink84, iben84}.  During the evolution of single-degenerate binaries, the system undergoes a nuclear burning phase, appearing as a SSS \citep{Rappaport94}. 
 The second pathway has recently been shown to also give rise to SSS characteristics, as pre-double-degenerate systems undergo a period in which the initial white dwarf accretes and burns matter from the companion's red giant wind \citep{distefano10ii}. However, for either case, observations show a lack of soft X-ray sources \citep{gilfanov10, distefano10i}.  The majority of observed SSS systems are now understood to be novae, in which previously accreted matter is burned on the WD surface \citep{Pietsch05}, but WDs which burn H in nova eruptions are thought to lose, not gain, mass  \citep{Nomoto82}.
 It is also clear that the majority of known Galactic non-nova supersoft X-ray sources do not contain red giants as donors, but rather main-sequence stars--with inferred donor masses smaller than expected \citep{Cowley98}.  Recently massive (Be?) companions have been inferred for some SSSs in M31 \citep{Orio10}.
 It remains possible that a large number of SSSs emit predominantly in the extreme UV, below \Chandra\ detection limits \citep{Southwell96,Hachisu96,Hachisu03}.  
This possibility makes observations of apparently transient SSSs critical to the understanding of Type Ia progenitors.

Currently five or six SSSs (besides novae) are known within our galaxy; however, only one has been seen within a globular cluster \citep{Greiner00}. Located within the M3 cluster, the transient supersoft X-ray source 1E1339.8+2837 (hereafter 1E1339) was first observed with the Einstein obervatory, possessing a 0.5-4.5 keV luminosity of $4\times10^{33}$ erg s $^{-1}$ \citep{Hertz83}. The {\it ROSAT} all-sky survey later detected a very soft X-ray source within the M3 globular cluster, with a maximum bolometric luminosity of $2.0\times10^{35}$ erg s $^{-1}$ \citep{Verbunt95}.  In 1992 January the {\it ROSAT} High-Resolution Imager (HRI) observed 1E1339 to have a soft spectrum ({\it kT} $\sim20$ eV) and a 0.1-2.4 keV luminosity of $2\times10^{35}$ erg s $^{-1}$; however, within six months the source had faded below HRI detection limits \citep{Hertz93}. During its peak, 1E1339 was roughly one to two orders of magnitude less luminous than an average SSS \citep[see][]{Kahabka97}. 

After the 1992 outburst, 1E1339 fell into a period of relative quiescence,
 adopting a hard X-ray spectrum more typical of cataclysmic variables (CVs) \citep[e.g.][]{Baskill05}.
A follow up {\it ASCA} observation taken in 1997 confirmed the luminosity decrease, and found the spectrum of 1E1339 to be best approximated by a 5 keV bremsstrahlung model, with a 0.5-10 keV luminosity of $1.6\times10^{33}$ erg s $^{-1}$ \citep{Dotani99}. The optical counterpart for 1E1339 was identified by \citet{Edmonds04} near the subgiant branch in $V-I$ color-magnitude diagrams, but to be extremely bright in the $U$ band, with strong variability.  Its X-ray to optical flux ratio ($F_X/F_{opt}\sim1$) is far too low for X-ray binaries containing neutron stars, 
high for normal CVs, but consistent with magnetic CVs \citep{Verbunt97}.  
The extremely blue colors of 1E1339's optical counterpart \citep{Edmonds04} allow us to infer a high rate of accretion, which combined with the relatively low-luminosity X-ray emission rules out a neutron star or black hole accretor; we regard a WD accretor as certain.  
The small inferred radius of the SSS episode, and the high $F_X/F_{opt}$ ratio, suggested a magnetic CV nature \citep{Edmonds04}.
In this paper we report the results of three \Chandra\ ACIS-S observations of 1E1339.

\section{Data Reduction}\label{s:obs}
We observed the X-ray source 1E1339 on three separate occasions with the Chandra ACIS-S, using three active CCDs (S2, S3 and S4) with a 1/2 subarray (to reduce pileup in case 1E1339 were to become extremely bright) using the ACIS {\it very faint} mode. 
ObsID 4542 was taken November 11 2003 for 10.4 ks, ObsID 4543 on May 9 2004 for 10.3 ks, and ObsID 4544 on January 10 2005 for 9.8 ks. 
In addition, we obtained HST observations of 1E1339 in several bands, from F814W ($I$) to F220W (near-UV), coincident with the 2004 Chandra
observation. An analysis of the HST observations will be published elsewhere.

\begin{figure}
\figurenum{1}
\includegraphics[scale=0.55]{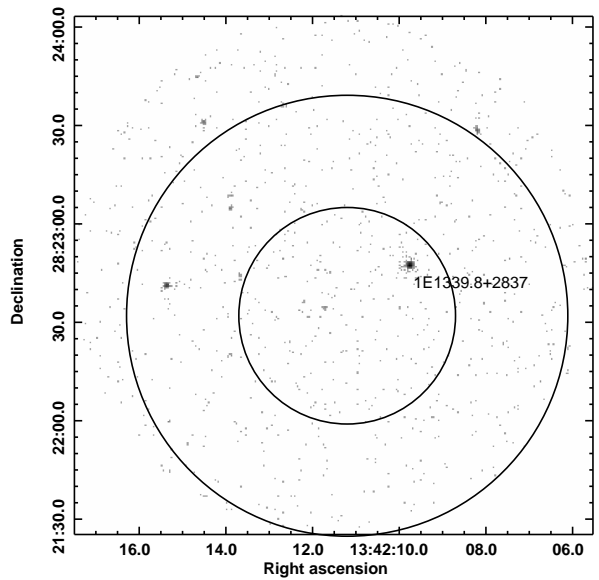}
\caption[obmap.ps]{ \label{fig:obmap}
Source map of X-ray sources within the M3 globular cluster (NGC 5272) from the merged data of the three observations. The core and half-mass radii are shown. 1E1339.8+2837 can be seen as the brightest X-ray source.
} 
\end{figure}

The data were reduced using CIAO version 4.2\footnote{http://cxc.harvard.edu/ciao}, as well as FTOOLS\footnote{http://heasarc.gsfc.nasa.gov/docs/software/ftools/ftools\_menu.html}. 
A bad pixel file was created with the CIAO {\it acis\_run\_hotpix} tool. The observations were then reprocessed (using {\it acis\_process\_events}) and further filtered\footnote{http://cxc.harvard.edu/ciao/threads/createL2/}
 to create new level 2 event files.
 In order to create a source map, the images upon the S3 chip were adjusted, then merged together. The merged image was then divided by the merged set of exposure maps, taken from each observation (see Figure \ref{fig:obmap}). The position of 1E1339 was determined to be R.A. = 13$^{h}$42$^{m}$09.8$^{s}$, Decl. = +28$^\circ$22\arcmin47\arcsec (J2000), in agreement with previous {\it HST} and {\it ROSAT} HRI observations \citep[see][]{Edmonds04}. We detected multiple other X-ray sources, which will be reported separately.

 From each individual observation, we selected a circle with a radius of 2\arcsec around 1E1339 and chose an annulus of radius 10\arcsec that excluded the source for the background. Using the task {\it psextract}, we generated RMF and ARF files for the source and background spectra. For each observation, an improved RMF file was created using the {\it mkacisrmf} task, over an energy range from 0.3 to 10 keV.

\subsection{X-ray Variability}
\begin{figure}
\figurenum{2}
\includegraphics[scale=0.6]{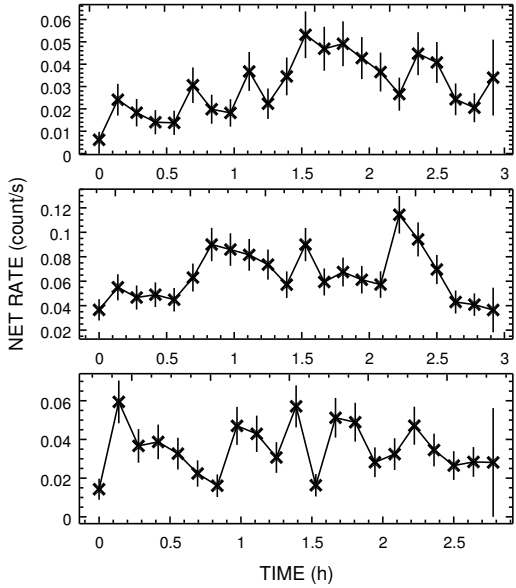}
\caption[multi2.ps]{ \label{fig:multi}
 Light curves extracted from the three Chandra observations (ObsID 4542 top, ObsID 4543 middle and ObsID 4544 bottom) over the observational period, using a binning of 500 s. Corresponding variability indexes of 7 (variable) for 4542, 6 (variable) for 4543 and zero (not variable) for 4544 were found using the CIAO {\it glvary} tool.
} 
\end{figure}

The three observations were found to have notable differences within their average count rates, ranging from a maximum of $6.6\times10^{-2}$ count s $^{-1}$ seen in the second observation (ObsID 4543) to a minimum of $3.0\times10^{-2}$ count s $^{-1}$ seen in the first observation (ObsID 4542, see Table \ref{tab:1339count}). 
We further extracted lightcurves from each of the three observations of 1E1339.8+2837 with use of the CIAO {\it dmextract} tool, using a binning of 500 s (see Figure \ref{fig:multi}).
We then tested the variability for each individual observation, making use of the {\it dither\_region} tool and the {\it glvary} tool, which applies the Gregory-Loredo algorithm. The algorithm divides the observation in various time bins and looks for significant deviations, reporting a variability index \citep{Gregory92}. We found the first and second observations to be variable (variability index of 7 and 6, corresponding to a variability probability of 0.998 and 0.989 respectively), while the third observation (ObsID 4544) showed no variability (variability index of zero, corresponding to a variability probability of 0.050).

\section{X-ray Spectral Analysis}\label{s:spec}
We extracted X-ray spectra from each of the three observations of 1E1339 and used XSPEC version 12.5 to model them. The spectra from the first and third observation (ObsID 4542 and 4544, respectively) were binned using the GRPPHA tool so that each bin contained a minimum of 15 counts. The second observation (ObsID 4543) contained significantly more counts and was grouped to contain a minimum of 20 counts per bin. The first two observations displayed hard spectra typical of CVs; however, the third observation contained a large low energy tail (63 counts between 0.2-0.3 keV). Neither of the other observations showed this tail, containing minimal counts over the same range.  The sensitivity of the ACIS-S detector, and the accuracy of the \Chandra\ response files, fall off at low energies (below ${\sim0.5}$ keV). This introduces difficulty when trying to fit a spectrum over this range. For this reason, we ignored all counts below 0.5 keV for our primary fits.

 We included a photoelectric absorption component within our X-ray analysis, using a hydrogen column density, $N_H$, frozen at $5.5\times10^{19}$ cm$^{-2}$, inferred \citep{Predehl95} from the measured color excess E(B-V)=0.01 \citep{Harris10}\footnote{Updated 2010; 
http://www.physwww.mcmaster.ca/${\sim}$harris/mwgc.dat}.  We used the convolution component {\it cflux} to calculate the flux over a range of 0.5 to 10 keV. Each spectrum was then fit to a hot X-ray plasma emission model (XSPEC model {\it mekal}). The metal abundance was fixed, using [Fe/H]=-1.57 \citep{Harris96}.

\begin{figure}
\figurenum{3}
\includegraphics[scale=0.35, angle=270]{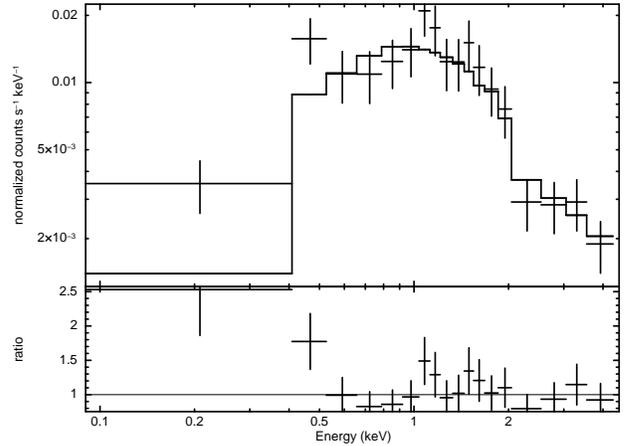}
\caption[4542fit.ps]{ \label{fig:4542}
Chandra spectrum of 1E1339.8+2837 (ObsID 4542) with an overplotted best-fit {\it mekal} model fit from 0.5-10 keV. The ratio of the data to the model is shown below. The fit yields $\chi^2_{\nu}$=0.47 and a null hypothesis probability=0.96.
} 
\end{figure}

\begin{figure}
\figurenum{5}
\includegraphics[scale=0.35, angle=270]{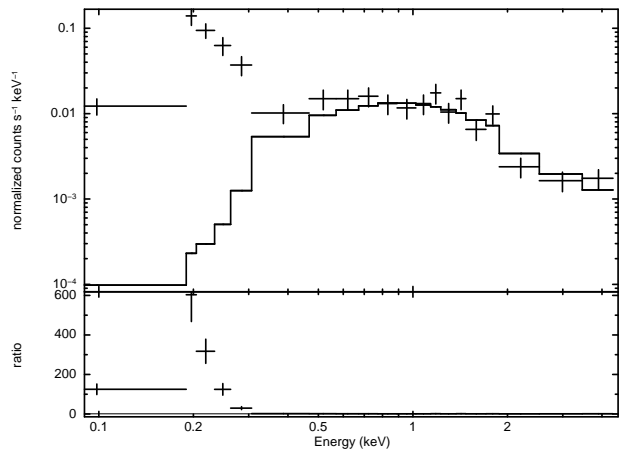}
\caption[4544fit.ps]{ \label{fig:4544}
Chandra spectrum of 1E1339.8+2837 (ObsID 4544) including the low energy tail, visible from 0.2-0.3 keV, with an overplotted best-fit {\it mekal} model fit from 0.5-10 keV (thus excluding the soft emission). The ratio of the data to the model is shown below. The fit yields $\chi^2_{\nu}$=1.09 and a null hypothesis probability=0.37.} 
\end{figure}

The three observations were fit with varying levels of success (see Table \ref{tab:1339fit}). From the hot diffuse gas emission model, the largest observed flux (0.5-10 keV) was $6.0\pm0.8\times10^{-13}$ erg cm$^{-2}$ s$^{-1}$, seen in the second observation, while the smallest flux was $1.8\pm0.4\times10^{-13}$ erg cm$^{-2}$ s$^{-1}$, seen in the third observation. Using a distance of 10.4 kpc to the M3 cluster \citep{Harris96}\footnotemark[4], these correspond to luminosities of $7.7\pm1.0\times10^{33}$ erg s $^{-1}$ and $2.4^{+0.6}_{-0.5}\times10^{33}$ erg s $^{-1}$. The individual fits can be seen in Figures \ref{fig:4542}-\ref{fig:4544}.  In addition, the observations were fit to an absorbed power-law spectrum. In the case of the first two observations this gave a worse fit; however, for the third observation this gave a slightly better fit and a marginally higher luminosity ($L_X=2.6\pm0.5\times10^{33}$ erg s $^{-1}$ over the 0.5-10 keV range, again see Table \ref{tab:1339fit}).
We also tested the models leaving the photoelectric absorption as a free parameter; however, the $N_H$ values were consistent with the cluster value, and an F-test showed insignificant improvement. Finally, we tested the models with a pileup component. This served to worsen the fit in all three observations and slightly altered the luminosities ($L_X=3.9^{+0.6}_{-1.0}\times10^{33}$ erg s $^{-1}$ for the first observation, $L_X=7.4^{+2.2}_{-1.0}\times10^{33}$ erg s $^{-1}$ for the second observation and $L_X=2.4\pm0.6\times10^{33}$ erg s $^{-1}$ for the third observation). 

\begin{figure}
\figurenum{4}
\includegraphics[scale=0.35, angle=270]{f4.eps}
\caption[4543fit.ps]{ \label{fig:4543}
Chandra spectrum of 1E1339.8+2837 (ObsID 4543) showing an overplotted best-fit {\it mekal} model fit from 0.5-10 keV. The ratio of the data to the model is shown below. The fit yields $\chi^2_{\nu}$=1.50 and a null hypothesis probability=$5.0\times10^{-2}$. 
} 
\end{figure}

We created light curves for each observation over the 0.5 to 10 keV range. Assuming a linear relation between the luminosity from the {\it mekal} models and the count rate, we used the maximum and minimum count rates from the modified light curves to estimate a maximum and minimum luminosity. 
From this we found an estimated maximum 0.5-10 keV luminosity of $1.4\pm0.3\times10^{34}$ erg s$^{-1}$, observed in the second observation, and a minimum luminosity of $8.5^{+4.9}_{-4.6}\times10^{32}$ erg s$^{-1}$, observed in the third observation, indicating that 1E1339's hard X-ray flux has varied by roughly an order of magnitude. 

We can estimate the mass transfer rate implied using the formulae of \citet{Patterson85}, converting to the 0.5-10 keV band using our power-law fit to get $L_X$(0.5-10 keV)$=1.6\times10^{32}\dot{M}^{1.0}_{16}M^{1.8}_{0.7}$ erg s$^{-1}$ where $M_{0.7}$ is the white dwarf mass in units of 0.7 \Msun\ and $\dot{M}_{16}$ is the mass transfer rate in units of $10^{16}$ g s$^{-1}$.  This gives a peak mass transfer rate of $2.5-8.8\times10^{17}$ g s$^{-1}$, or $3.7-13.3\times10^{-9}$ \Msun/yr (for WD masses of 0.7 to 1.4 \Msun).  
Such high mass transfer rates are typically seen in novalike CVs, or during the outbursts of dwarf nova CVs.  However, the hard X-ray flux of such systems is typically much smaller \citep{Patterson85, Wheatley96}.
The CVs with the highest hard X-ray luminosities are typically magnetic systems \citep{Verbunt97}, since matter flows onto the white dwarf in a stream following the magnetic field (rather than a disk) and loses half of its gravitational potential energy in an optically thin post-shock region \citep{Cropper99} rather than an accretion disk which becomes optically thick \citep{Patterson85}.  Thus, the peak hard X-ray luminosity of 1E1339 can be taken as evidence in favor of the magnetic nature of the accretion flow, as suggested by \citet{Edmonds04}.

\subsection{Low-Energy Emission}
The low-energy emission seen in ObsID 4544 does not seem to be due to a calibration problem, since such very soft X-rays are not seen elsewhere on the detector and the optical counterpart is far below the $V$ magnitude ($\sim$8.1) required for optical leakage.\footnote{http://cxc.harvard.edu/proposer/POG/html/chap6.html\#tth\_sEc6.3}  
Some excess low-energy emission may also be seen in ObsIDs 4542 and 4543 (see Figures \ref{fig:4542} and \ref{fig:4543}).  For ObsID 4544, we attempted to model the low energy emission by adding a blackbody component (XSPEC model {\it bbodyrad}) to the original hot diffuse gas emission model and including data in the fit down to 0.2 keV. The best fit resulted in a plasma temperature 
 of $\sim7$ keV, a blackbody temperature of $\sim10$ eV and a 0.2-10 keV luminosity of $5.0\times10^{33}$ erg s $^{-1}$. However, the fit produced a poor $\chi^2_{\nu}$ of 4.41.  We are unable to fit with {\it any} model the energy distribution of the detected counts using the current ACIS response matrices, which are known to poorly model the ACIS response below 0.3 keV. 
We may guess that the bolometric luminosity of this low-energy component (mostly in the extreme UV) may exceed the luminosity of the hard component, although we cannot prove this. 

Significant blackbody-like luminosity from the boundary layer has long been predicted from CVs \citep[e.g.][]{Patterson85b}.  Such blackbody-like luminosity has only been seen in the X-ray regime in magnetic CVs, such as polars \citep[e.g.][]{Ramsay04}.  
1E1339's high extreme UV luminosity is reminiscent of the high far-UV luminosity of the CV AKO9 in 47 Tuc, attributed to a high rate of mass transfer through its accretion disk \citep{Knigge03}.  AKO9 also has a subgiant donor, and its blue far-UV spectrum and presence of He {\tt II} $\lambda$4686 emission suggest a luminous ($\sim10^{34}$ ergs s$^{-1}$) EUV-bright boundary layer \citep{Knigge03}.  
However, if 1E1339's soft emission is more luminous than its hard emission (which will require observations with better low-energy calibration to determine), then new models may be required.  
It is possible, for instance, that the very soft emission arises from the same (uncertain) processes that resulted in the 1991-1992 supersoft state, as the luminosity may be similar.   
If accretion wind models \citep{Hachisu03} are able to shift the photosphere temperature of SSSs in similar ways, they may explain how Type 1a supernova progenitors could hide from X-ray detection \citep{gilfanov10, distefano10i,distefano10ii}.  We note that it is unclear how an expanding photosphere driven by an accretion wind could coexist with the luminous, unabsorbed hard X-ray emission seen in 1E1339.

\section{Comparison to Other Results}

\subsection{Comparison to Previous Observations}

During the 1992 outburst, 1E1339 appeared as a supersoft X-ray source, with a 0.1-2.4 keV luminosity of $2 \times10^{35}$ erg s $^{-1}$ \citep{Hertz93}. Following this, the source fell into a period of relative quiescence. 
In 1997, a follow up ASCA observation found 1E1339 to possess a CV-type spectrum, as well as a 0.5-10 keV luminosity of  $1.6\times10^{33}$ erg s $^{-1}$ \citep{Dotani99}.
Our inferred maximum luminosity of $1.4\pm0.3\times10^{34}$ erg s $^{-1}$ still lies considerably below that of the 1992 outburst; however, it greatly exceeds the 1997 observation. Our minimum inferred 0.5-10 keV luminosity of $8.5^{+4.9}_{-4.6}\times10^{32}$ erg s $^{-1}$ is seen to be considerably lower than the 1997 observation, again highlighting the variable nature of 1E1339.

\subsection{Comparison to GK Persei}

GK Persei is generally considered to be the upper limit of observed hard X-ray luminosity for CVs. The system is an intermediate polar with 
an unusually long orbital period of 2 days \citep{Crampton86} and a short white dwarf period of 351 s \citep{Watson85}. Roughly every three years, the system undergoes outbursts \citep{Simon02} during which time it reachs an average 2-10 keV flux of $\sim2.5\times10^{-10}$ erg cm$^{-2}$ s$^{-1}$ \citep{Evans09}. 

Our maximum observed luminosity for 1E1339 ($1.4\pm0.3\times10^{34}$ erg s $^{-1}$ over 0.5-10 keV) shows 1E1339 to be exceedingly luminous for a CV, perhaps more luminous than GK Persei. Observations of a recent outburst from GK Persei in 2006 give a maximum 2-10 keV flux of $3.3\times10^{-10}$ erg cm$^{-2}$ s$^{-1}$ 
However, an uncertainty in the distance to GK Persei, ranging from ${\sim340}$ pc \citep{Warner87} to 525 pc \citep{Duerbeck81}, allows for a wide range of luminosities for the 2006 outburst, from $4.6\times10^{33}$ erg s$^{-1}$ to $1.1\times10^{34}$ erg s$^{-1}$. To compare the luminosity of 1E1339 to GK Persei, we made use of the \Chandra\ PIMMS version 3.9k to estimate our observed luminosity within the 2-10 keV range. Using a thermal bremsstrahlung model with an energy of 22 keV, we found a corresponding 2-10 keV luminosity of $9.8\pm1.4\times10^{33}$ erg s$^{-1}$, while a power law model with photon index of 1.3 resulted in a luminosity of $1.12\pm0.15\times10^{34}$ erg s$^{-1}$. Both the temperature and photon index are taken from the best-fit models. Depending upon the true distance to GK Per, 1E1339 may exceed GK Persei in X-ray luminosity, making 1E1339 the CV with the highest hard X-ray luminosity. As estimates for the distance to GK Persei cover a large range, 1E1339 provides a more accurate estimate of the maximum observed CV hard X-ray luminosity.

\subsection{Comparison to Very Faint X-ray Transients}
The maximum luminosity falls into the range of very faint X-ray transients (VFXTs), with peak luminosities $10^{34}$-$10^{36}$ erg s$^{-1}$ \citep{Wijnands06}, and is comparable to previously studied VFXTs \citep[see][]{Muno05}. Several explanations have been proposed for VFXTs, including low-mass X-ray binaries (LMXBs) with extremely low-mass companions  \citep{King06},  and classical and recurrent nova outbursts \citep{Mukai08}. The large peak X-ray luminosity of 1E1339 suggests that the most luminous CV-type systems may also account for some non-nova VFXTs.

\section{Conclusions}
From three observations of the X-ray source 1E1339, we found the source to be considerably variable, with a 0.5-10 keV luminosity ranging from $1.4\pm0.3\times10^{34}$ erg s$^{-1}$ to $8.5^{+4.9}_{-4.6}\times10^{32}$ erg s$^{-1}$. The maximum observed X-ray luminosity gives 1E1339 the largest well-measured hard X-ray luminosity for any CV, providing a reliable estimate on the X-ray luminosity achievable by CV-type systems.  The peak X-ray luminosity falls into the range of VFXTs, suggesting that the most X-ray luminous CV-type systems may account for some VFXTs.

Above 0.5 keV, we were able to fit each observation to a hot diffuse gas emission model with varying levels of success, showing a typical hard CV spectrum. Our observations over 0.5-10 keV are in agreement with previous observations which have shown 1E1339 to have fallen into a period of relative quiescence after its 1992 SSS period \citep[see][]{Dotani99}. However, a luminous low energy component, below 0.3 keV, was seen in the 2005 observation. We were unable to successfully model this component of the observation, due to the uncertain calibration of the ACIS-S detector below 0.3 keV. However, it is possible that the bolometric luminosity of the low energy component may exceed that of the hard component.

1E1339's unusual behavior raises a number of questions.  What is the origin and luminosity of the luminous low energy component?  
Is this (largely hidden) low-energy flux due to nuclear burning on (part of?) the white dwarf surface \citep{Edmonds04}, or to a blackbody-like accretion-powered component, like those seen in magnetic CVs?   What is the geometry of the accretion flow?  (Can this be tested with longer, continuous X-ray observations?)  Was the 1992 outburst an isolated nuclear flash, an increase in the accretion flow, or a change in the photospheric radius?  
Our near-simultaneous near-UV HST observations may help address these questions. There is also a clear need for XMM observations, as the pn camera has better-calibrated responses down to 0.2 keV to characterize the soft X-ray emission, and a long XMM observation could search for pulsations and a WD spin period, expected for a magnetic CV.  
Understanding the behavior of low energy components such as that reported here may play a critical role in understanding the mismatch between observed and predicted numbers of SSSs reported by \citet{gilfanov10} and \citet{distefano10i}.

\begin{deluxetable*}{ccccccc}
\tablewidth{5truein}
\tablecaption{\textbf{Count rates for observations of 1E1339.8+2837}}
\tablehead{
& & & \multicolumn{2} { c }{Count Rate} \\
\colhead{\textbf{Date}}  & ObsID & Exp. Time  & 0.5-10 keV & $<$ 0.3 keV  \\
}

\startdata
2003 Nov 11 & 4542 & 10.4 &  $2.7^{+0.2}_{-0.2}\times10^{-2}$ &$8.6^{+5.0}_{-3.4}\times10^{-4} $    \\ 
2004 May 9 & 4543 & 10.3 &  $6.1^{+0.3}_{-0.3}\times10^{-2}$ &$1.7^{+0.7}_{-0.5}\times10^{-3}   $  \\ 
2005 Jan 10 & 4544 & 9.8 & $2.3^{+0.2}_{-0.2}\times10^{-2}$ &$1.0^{+0.1}_{-0.1}\times10^{-2}    $ \\ 
\enddata
\tablecomments{Count rates for observations of 1E1339.8+2837. A notable low energy emission is seen in the 2005 observation. Errors are 90\% confidence.
Exposure times in ks.
All count rates in counts s $^{-1}$.}
\label{tab:1339count}
\end{deluxetable*}

\begin{deluxetable*}{ccccccc}
\tablewidth{5truein}
\tablecaption{\textbf{Spectral Fits to 1E1339.8+2837}}
\tablehead{
\colhead{\textbf{Epoch}}  & $N_H\times10^{19}$ & $\Gamma$/kT  & $L_X$ & $\chi^2_{\nu}$/dof  & nhp  \\
}
\startdata
\multicolumn{6}{c}{PHABS * MEKAL Fit} \\
\hline \\
2003 & (5.5) & $>$11.4 & 3.8$^{+0.5}_{-0.8}\times10^{33}$ & 0.47/15 & 0.956   \\ 
2004 & ... & $22^{+55}_{-10}$ & 7.7$^{+1.0}_{-1.0}\times10^{33}$ & 1.50/26 & 0.050 \\ 
2005 & ... & $8^{+17}_{-4}$ & 2.4$^{+0.6}_{-0.5}\times10^{33}$ & 1.09/12  & 0.365\\  
\hline \\
\multicolumn{6}{c}{PHABS * Power-law Fit} \\
\hline \\
2003 & (5.5) & $1.3^{+0.2}_{-0.2}$ & 3.9$^{+0.7}_{-0.7}\times10^{33}$ & 0.49/15 & 0.946   \\ 
2004 & ... & $1.3^{+0.1}_{-0.1}$ & 8.3$^{+1.0}_{-0.9}\times10^{33}$ & 1.56/26 & 0.035 \\ 
2005 & ... & $1.4^{+0.2}_{-0.2}$ & 2.6$^{+0.5}_{-0.5}\times10^{33}$ & 0.89/12 & 0.557 \\ 
\enddata
\tablecomments{Spectral fits to 1E1339.8+2837 data from 2003, 2004 and 2005 (ObsID 4542, 4543 and 4544, respectively) to either a PHABS * MEKAL model or PHABS * Power-law model.  
The hydrogen column density ($N_H$) was held fixed.
$L_X$ in erg s$^{-1}$ for 0.5-10 keV.
The last two columns list the reduced chi-squared value, the degrees of freedom, and the null-hypothesis probability. 
\label{tab:1339fit}}
\end{deluxetable*}


\acknowledgements

\bibliographystyle{apj}

\bibliography{M3bib}

\end{document}